\documentclass[aps,twocolumn,showpacs,preprintnumbers,floatfix,nofootinbib]{revtex4-1}
\usepackage{graphicx}
\usepackage{color}      
\usepackage[subfigure]{graphfig}
\usepackage{epsfig}
\usepackage{dcolumn}
\usepackage{amsmath}
\usepackage[
             colorlinks=true,%
             linkcolor=blue,%
             citecolor=blue
             ]{hyperref}

\def\be{\begin{equation}}
\def\ee{\end{equation}}

\definecolor{green}{rgb}{0,.5,0}
\definecolor{red}{rgb}{1,.0,0}

\immediate\write18{texcount -inc -incbib 
-sum borra.tex > /tmp/wordcount.tex}

\begin{document}

\title{\vspace{1.0in} {\bf $\pi$N and strangeness sigma terms at the physical point with chiral fermions}}


\author{{Yi-Bo Yang$^{1}$, Andrei Alexandru$^{2}$,  Terrence Draper$^{1}$, Jian Liang$^{1}$, and Keh-Fei Liu$^{1}$}
\vspace*{-0.5cm}
\begin{center}
\large{
\vspace*{0.4cm}
\includegraphics[scale=0.20]{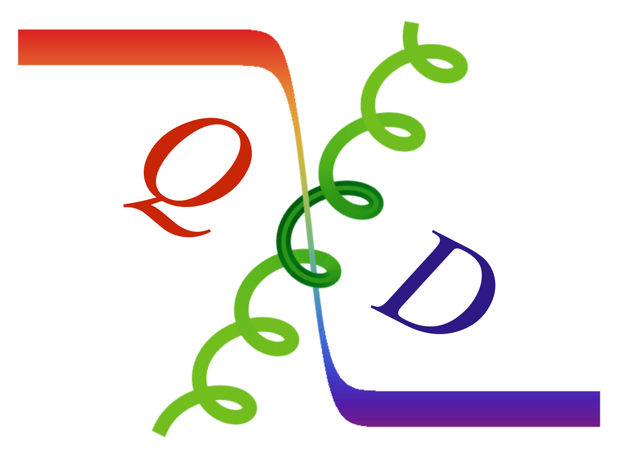}\\
\vspace*{0.4cm}
($\chi$QCD Collaboration)
}
\end{center}
}
\affiliation{
$^{1}$\mbox{Department of Physics and Astronomy, University of Kentucky, Lexington, KY 40506, USA}\\
$^{2}$\mbox{Department of Physics, The George Washington University, Washington, DC 20052, USA}
}

\begin{abstract}
Lattice QCD calculations with chiral fermions of the $\pi$N sigma term $\sigma_{\pi N}$  and  strangeness sigma term $\sigma_{sN}$ 
including chiral interpolation with continuum and volume corrections are provided in this work, with the excited-state contaminations subtracted properly. We calculate the scalar matrix element for the light/strange quark directly and find $\sigma_{\pi N}=45.9(7.4)(2.8)$ MeV, with the disconnected insertion part contributing 20(12)(4)\%, and  $\sigma_{sN}=40.2(11.7)(3.5)$ MeV, which is somewhat smaller than $\sigma_{\pi N}$. The ratio of the strange/light scalar matrix elements is $y$ = 0.09(3)(1).
\end{abstract}


\maketitle

\section{Introduction}

 The $\pi$N sigma term $\sigma_{\pi N}$ for the light quark is defined as
\begin{eqnarray}
\sigma_{\pi N}\equiv \hat{m}\langle N|\bar{u}u+\bar{d}d|N\rangle,
\end{eqnarray}
where $\hat{m}=(m_u+m_d)/2$ is the averaged light quark mass, $|N\rangle$ represents the nucleon state which is normalized as $\langle N|N\rangle = L^3$ in this case for the unpolarized nucleon at rest, and $\bar{u}u$ and $\bar{d}d$ are the quark bilinear operators. The strangeness sigma term $\sigma_{sN}$ is similarly defined with
$f_s^N$ being its fraction of the nucleon mass 
\begin{eqnarray}
\sigma_{sN} \equiv m_s\langle N|\bar{s}s|N\rangle,  \,\,\,\,\,\,\, 
f_s^N = \frac{\sigma_{sN}}{m_N}.
\end{eqnarray}

As measures of both explicit and spontaneous chiral symmetry breakings in the baryon sector, $\sigma_{\pi N}$ and $\sigma_{sN}$ are fundamental quantities  which pertain to a wide range of issues in hadron physics, such as the quark mass contribution in the baryon which is related to the Higgs coupling to the observable matter~\cite{Gong:2013vja,Young:2009zb,Shanahan:2012wh}, the pattern of SU(3)
breaking~\cite{Young:2009zb}, $\pi N$ and $KN$ scatterings~\cite{Brown:1971pn,Cheng:1972pe}, and kaon condensate in dense matter~\cite{Kaplan:1986yq}. Using a sum rule for the nucleon mass, the heavy quark mass contribution can be deduced from that of the light flavors,  in the leading order of the strong coupling and the heavy quark limit~\cite{Shifman:1978zn,Gong:2013vja}. At the same time, precise values of the quark mass term for various flavors, from light to heavy, are of significant interest for dark matter searches~\cite{Falk:1998xj,Ellis:2008hf,Giedt:2009mr}, where the popular candidates for dark matter (such as the weakly interacting massive particle) interact with the observable world through the Higgs couplings, so that the precise determination of the $\sigma_{\pi N}$ and $\sigma_{sN}$ can provide remarkable constraints on the direct detection of the dark matter candidates.

Phenomenologically,  the $\sigma_{\pi N}$ term is typically extracted from the $\pi N$ scattering amplitude. To lowest order in $m_{\pi}^2$, 
the unphysical on-shell isospin-even $\pi N$ scattering amplitude at the Cheng-Dashen point corresponds to 
$\sigma(q^2=2m_{\pi}^2)$~\cite{Brown:1971pn,Cheng:1972pe} which can be determined from $\pi N$ scattering via fixed-$q^2$ dispersion 
relation~\cite{Cheng:1972pe}. $\sigma_{\pi N}$ at $q^2 = 0$ can be extracted through a soft correlated two-pion form 
factor~\cite{Gasser:1990ce,Becher:1999he,Pavan:2001wz}. Also, baryon chiral perturbation theory and Cheng-Dashen theorem have been used to analyze the $\pi N$ scattering amplitude for $\sigma_{\pi N}(0)$. They give $\sigma_{\pi N}$ values in the range
$\sim 45 - 64$ MeV, while the most recent analysis~\cite{Hoferichter:2015dsa} gives 59.1(3.5) MeV.

Both $\sigma_{\pi N}$ and $\sigma_{sN}$ are amenable to lattice QCD calculations and there are two ways to calculate them. One
is via the Feynman-Hellman {theorem} and the other is by directly calculating the matrix elements through the ratio of 3-pt and
2-pt correlation functions.

Following the Feynman-Hellman theorem (FH)
\begin{eqnarray}\label{eq:FHT}
\sigma_{\pi N} = m_q \frac{\partial m_N(m_q)}{\partial m_q} |_{m_q=\hat{m}^{phys}}
\end{eqnarray}
where $\hat{m}^{phys}$ is the quark mass corresponding to the physical $m_{\pi}$, one can calculate the nucleon mass at different quark masses and obtain 
$\sigma_{\pi N}$.
A number of such calculations have been performed~\cite{Ohki:2008ge,Durr:2011mp,Horsley:2011wr,Alexandrou:2014sha,Bali:2012qs,Durr:2015dna}, and analyses with chiral extrapolation based on lattice data have also been carried out~\cite{Young:2009zb,Shanahan:2012wh,Alvarez-Ruso:2013fza,Lutz:2014oxa,Ren:2014vea}. 
Similarly, there have also been a number of direct calculations of $\sigma_{\pi N}$ scalar matrix elements (ME) over the 
years~\cite{Dong:1995ec,Fukugita:1994ba,Bali:2011ks,Dinter:2012tt,Abdel-Rehim:2016won,Bali:2016lvx}, that use Wilson-type fermions which explicitly break chiral symmetry. The most recent three lattice calculations obtained consistent results regardless of whether with the FH theorem~\cite{Durr:2015dna} or direct matrix element calculation~\cite{Abdel-Rehim:2016won,Bali:2016lvx}, but the common value is around 37(4) MeV and is almost 5$\sigma$ smaller than the recent phenomenological analysis~\cite{Hoferichter:2015dsa} mentioned above.

Before investigating other avenues to understand the tension between the lattice simulation and phenomenological analysis~\cite{Hoferichter:2016ocj}, a question that cannot be avoided is whether the explicit breaking of chiral symmetry by lattice artifacts, as in the case of Wilson-type fermions, is responsible for the difference. Due to explicit chiral symmetry
breaking, the quark mass has an additive renormalization and the flavor-singlet and non-singlet quark masses renormalize differently. As a consequence, the strangeness content can be mixed with those of $u$ and $d$~\cite{Takeda:2010cw,Bali:2011ks} leading to a larger value. Attempts have been made to take the flavor-mixing into account which reduce $\sigma_{sN}$~\cite{Michael:2001bv,Bali:2011ks,Bali:2016lvx}, with the renormalization factors of the singlet and iso-vector part of the scalar ME differing by as much as 40\%~\cite{Bali:2011ks}.

In contrast, simulations with overlap fermion for the valence quarks have exact chiral symmetry at finite lattice spacing, and thus they are free of the flavor-mixing problem which afflicts Wilson-type fermions. But the inversion of the chiral fermion is known to be one magnitude more expensive than a Wilson-type fermion, which makes the approach numerically challenging. The major task of this work is overcoming the numerical difficulty to address the role of chiral symmetry for this quantity.  The properties of the overlap action allow us to apply the multi-mass algorithm to calculate a number of quark masses ranging from the light $u/d$ quark to the strange with little overhead
(compared to inversion with one mass). This, together with the use of the low-mode substitution (LMS) technique described in Ref.~\cite{Yang:2015zja,Gong:2013vja}, allows us to obtain hundreds of measurements with just a few inversions, thus overcoming the expensive cost of the overlap action required to obtain precise results.

We also present here a direct ME calculation of  $\sigma_{sN}$ without any systematic uncertainty about the flavor mixing of a Wilson-type fermion. As in the case of $\sigma_{\pi N}$, one can take the derivative of the proton mass with respect to the strange quark mass in the sea to get $\sigma_{sN}$. But both the calculations based on the FH theorem~\cite{Toussaint:2009pz,Junnarkar:2013ac,Durr:2015dna} and phenomenological determinations~\cite{Gasser:1990ce,Young:2009zb,Shanahan:2012wh,Alarcon:2012nr,Ren:2014vea} are not very precise since the strange quark dependence of the proton mass is very weak. On the other hand, there are several calculations which use the direct ME calculation~\cite{Abdel-Rehim:2016won,Bali:2016lvx,Takeda:2010cw,Bali:2011ks,Oksuzian:2012rzb,Dinter:2012tt,Engelhardt:2012gd,Gong:2013vja,Abdel-Rehim:2016won,Bali:2016lvx}. 
The present work is the first direct ME calculation with chiral fermions on 2+1 flavor configurations where the pion mass is at the physical point.

In addition to $\sigma_{\pi N}$ and $\sigma_{sN}$, the renormalization independent ratio often quoted in the literature,
\begin{eqnarray}
y=\frac{2\langle N|\bar{s}s|N\rangle}{\langle N|\bar{u}u+\bar{d}d|N\rangle},
\end{eqnarray}
can be obtained and it is useful to delineate the $SU(3)$ breaking pattern in the octect baryon spectrum. Its value has not been
well determined and the estimates change over time, reflecting the range of uncertainties of $\sigma_{\pi N}$ and $\sigma_{sN}$.

Since a precise value of $\sigma_{s N}$ is hard to obtain from the FH theorem approach and we want to present both $\sigma_{\pi N}$ and $\sigma_{sN}$ within the same framework to access the correlation between them,  we choose to use the direct ME calculation for both $\sigma_{\pi N}$ and $\sigma_{sN}$ to obtain the final predictions.

The numerical setup and the details are described in Sec. \ref{sec:num}. Sec. \ref{sec:res} provides our simulation results of $\sigma_{\pi N}$ and $\sigma_{sN}$, as well as a comparison with the results from phenomenological analyses and other lattice calculations. This article will be closed by a short summary in Sec. \ref{sec:sum}.

\section{Numerical\ setup}\label{sec:num}

In this work, we use the valence overlap fermions on $2 +1$ flavor domain-wall fermion (DWF) configurations~\cite{Aoki:2010dy} to carry out the calculation.  
The effective quark propagator of the massive
overlap fermion is the inverse of the operator $(D_c + m)$~\cite{Chiu:1998eu,Liu:2002qu}, where  $D_c$ is chiral, i.e. $\{D_c, \gamma_5\} = 0$~\cite{Chiu:1998gp} and its detailed definition can be found in our previous works~\cite{Li:2010pw,Gong:2013vja,Yang:2015zja}.
Numerical details regarding the calculation of the overlap operator, eigenmode deflation in inversion of the quark matrix, and the
$Z(3)$ grid smeared source with LMS to increase statistics are given in~\cite{Li:2010pw,Gong:2013vja,Yang:2015zja}.

\begin{table}[htbp]
\begin{center}
\caption{\label{table:r0} The parameters for the RBC/UKQCD configurations\cite{Blum:2014tka}: spatial/temporal size, lattice spacing, residual mass of the DWF sea, the sea strange quark mass under $\overline{MS}$ scheme at {2 GeV}, the pion mass with the degenerate light sea quark (both in unit of MeV), and the number of configurations used in this work.}
\begin{tabular}{ccccccc}
Symbol & $L^3\times T$  &a (fm) & $m_{res}^{(s)}a$ &$m_s^{(s)}$&  {$m_{\pi}$}   & $N_{cfg} $ \\
\hline
24I & $24^3\times 64$& 0.1105(3) & 0.00315(4)&120   &330  & 203  \\
32I &$32^3\times 64$& 0.0828(3) &0.00067(1)& 110   &300 & 309  \\
48I &$48^3\times 96$& 0.1141(2) & 0.00061(1) & 94.9   &139 & 81  \\
\hline
\end{tabular}
\end{center}
\end{table}

The $2+1$ flavor RBC/UKQCD DWF gauge configurations used are on $24^3\times64$ (24I),  $32^3\times64$ (32I)~\cite{Aoki:2010dy} and $48^3\times96$ (48I)~\cite{Blum:2014tka} lattices. Other parameters of the ensembles used are listed in Table~\ref{table:r0}. We used 5 quark {masses} from the range $m_{\pi}\in$(250, 400) MeV on the first two ensembles, {and 8} quark 
masses from $m_{\pi}\in$(114, 400) MeV on the last ensemble which has larger volume and thus allows a lighter pion mass {with} 
the constraint $m_{\pi}L>3$. 

Both the connected and disconnected insertions (CI/DI) contribute to the light quark contents, while the strange sigma term just comes from the disconnected insertion.

The scalar matrix elements are obtained from the ratio of the three-point function to the two-point function
\begin{eqnarray}   \label{eq:ratio}
R(t_f,t)&=&\frac{\langle 0|\int d^3 y\Gamma^e{\chi}(\vec{y},t_f){\cal O}(t)\sum_{\vec{x}\in G}\bar{\chi}_S(\vec{x},0)|0 \rangle}{\langle 0|\int d^3 y\Gamma^e\chi(\vec{y},t_f)\sum_{\vec{x}\in G}\bar{\chi}_S(\vec{x},0)|0 \rangle},
\end{eqnarray}
where $\chi$ is the standard proton interpolation field and $\bar{\chi}_S$ is the field with gaussian smearing applied to all three quarks. All the correlation functions from the source points $\vec{x}$ in the grid $G$ are combined to improve the the signal-to-noise ratio (SNR).  ${\cal O}(t)$ is the scalar current $\int d^3x{ \overline{\psi}_f(x,t){\psi}_f(x,t)}$ located at time slice $t$ and $\Gamma^e=\frac{1}{2}(1+\gamma_4)$.
When $t_f$ is large enough, $R(t_f,t)$ is equal to the bare scalar matrix element
\begin{eqnarray}
g_S&\equiv&\frac{\textrm{Tr}[\Gamma^e\langle P |\int d^3x{ \overline{\psi}_f(x)\psi_f(x)} |P \rangle]}{\textrm{Tr}[\Gamma^e\langle P|P \rangle]},
\end{eqnarray}
which is $t$ independent, plus $t$-dependent corrections,
\begin{eqnarray}\label{eq:two_term}
R(t_f,t) ={g_S}+C_1 e^{-\Delta E(t_f-t)}+ C_2 e^{-\Delta E t}+ C_3 e^{-\Delta Et_f},\nonumber\\
\end{eqnarray}
where $\Delta E$ is the energy difference between the ground state and the first excited state and $C_{1,2,3}$ are the combinations of weights involving the excited states. Then the $g_S$ we want to extract corresponds to the case $0\ll t \ll t_f$.

For each quark mass on each ensemble, we constructed the ratio $R(t_f,t)$ for three sink-source separations $t_f$ from 0.9 fm to 1.4 fm, and for all the current insertion times $t$ between the source and sink.

\subsection{Connected and Disconnected insertion}
For the connected insertion, we use the stochastic sandwich method (SSM) with low-mode substitution (LMS)~{\cite{Yang:2015zja}} to improve the SNR of the calculation. The stochastic sandwich method uses the stochastic source at the sink time slices to avoid repeating the production of the sequential propagators for different sinks and hadron states, but the final SNR is sensitive to that additional stochastic noise. Our improved stochastic method replaces the long-distance part of the stochastic propagator from the sink to the current by its all-to-all version, using the low-lying eigensystem of $D_c$, which suppresses the influence of the stochastic noise on the sink propagator.

A regular grid with 2 smeared sources in each spatial direction for the 24I and 32I lattices (4 for the 48I lattice) are placed
on 2 time slices for the 24I and 32I lattices (3 for the 48I lattice). The separation between the centers of the neighboring grids in the same time slice is $\sim 1.3$ fm and each smeared source has a radius of $\sim$ 0.5 fm.
On the sink side, several noise point-grid sources are placed at three slices $t_f$ which are $0.9-1.4$ fm away from the source time slices. Furthermore, the matrix elements of the light scalar contents are dominated by the low-mode part of $D_c$ so that the use of LMS on the propagators from the current to the sink notably reduces the number of noise propagators (from $t_f$) needed. More details of the stochastic sandwich method with low-mode substitution are given in Ref.~\cite{Yang:2015zja}.

\begin{table}[htbp]
\begin{center}
\caption{\label{table:setup} The source/sink setup on the ensembles, for the connected insertion. $N_{grid}$ is the pattern of the smeared points on a grid source with noises and $N_{src}$ is the number of the propagators with such a kind of source. Three sets of the pair ($\Delta_t^{i},N^i_{sink}$) are for the sink propagators, with $\Delta_t^i$ being the physical distance between the source and sink and $N^i_{sink}$  the number of noises with such a $\Delta_t^i$.}
\begin{tabular}{cccccccc}
Ensemble & $N_{grid}$ &  $N_{src}$ 
&$\begin{array}{c}\Delta_t^1(\textrm{fm}), \\N^1_{sink}\end{array}$
& $\begin{array}{c}\Delta_t^2(\textrm{fm}), \\N^2_{sink}\end{array}$
&  $\begin{array}{c}\Delta_t^3(\textrm{fm}), \\N^3_{sink}\end{array}$ \\
\hline
24I & $2^3\times 2$&1 &  0.88, 5   & 1.11, 5 & 1.33, 5    \\
32I &$2^3\times 2$&1 &  0.99, 3   & 1.16, 3 & 1.24, 3\\
48I &$4^3\times 3$&5 &  0.91, 4 & 1.14, 8 &1.37, 12 \\
\hline
\end{tabular}
\end{center}
\end{table}

The simulation setup for the connected insertion on three ensembles is listed in Table~\ref{table:setup}. We note that although the 48I ensemble has {a} larger volume which {can accommodate} more smeared grid points for 
 the source which improves the SNR with a single inversion, the SNR around the physical point is still small. So we used 5 propagators at the source to improve the SNR on the 48I by a factor of 2. 
The total cost on the 48I ensemble dominates and can be estimated by 34 inversions (with residual $10^{-7}$) per configuration. (The cost of the contraction with LMS is about 1 inversion per source propagator.)  The number of ideal equivalent measurements $N_{meas}$ for the grid source at the physical point on the 48I ensemble is 192 (points in grid) $\times$ 5 (sources)= 960 per configuration and 77,760 in total for the connected insertion.

The same noise grid-smeared sources are used in the production of the nucleon propagator for the disconnected insertion, and we loop over all the time slices for the nucleon source. The position of the grid is randomly shifted on each time slice.
As has been carried out
in previous studies of the strangeness content~\cite{Gong:2013vja} and quark spin~\cite{Gong:2015iir}, the quark loop is calculated 
with the exact low eigenmodes (low-mode average (LMA)) while the high modes are estimated with 8 sets of $Z_4$ noise on the same (4,4,4,2) grid with odd-even dilution and additional dilution in time. The vacuum expectation value of the quark loops has been subtracted before combining with the proton propagator to get the correlated three-point function.

\begin{figure}[h]
  \includegraphics*[scale=0.7]{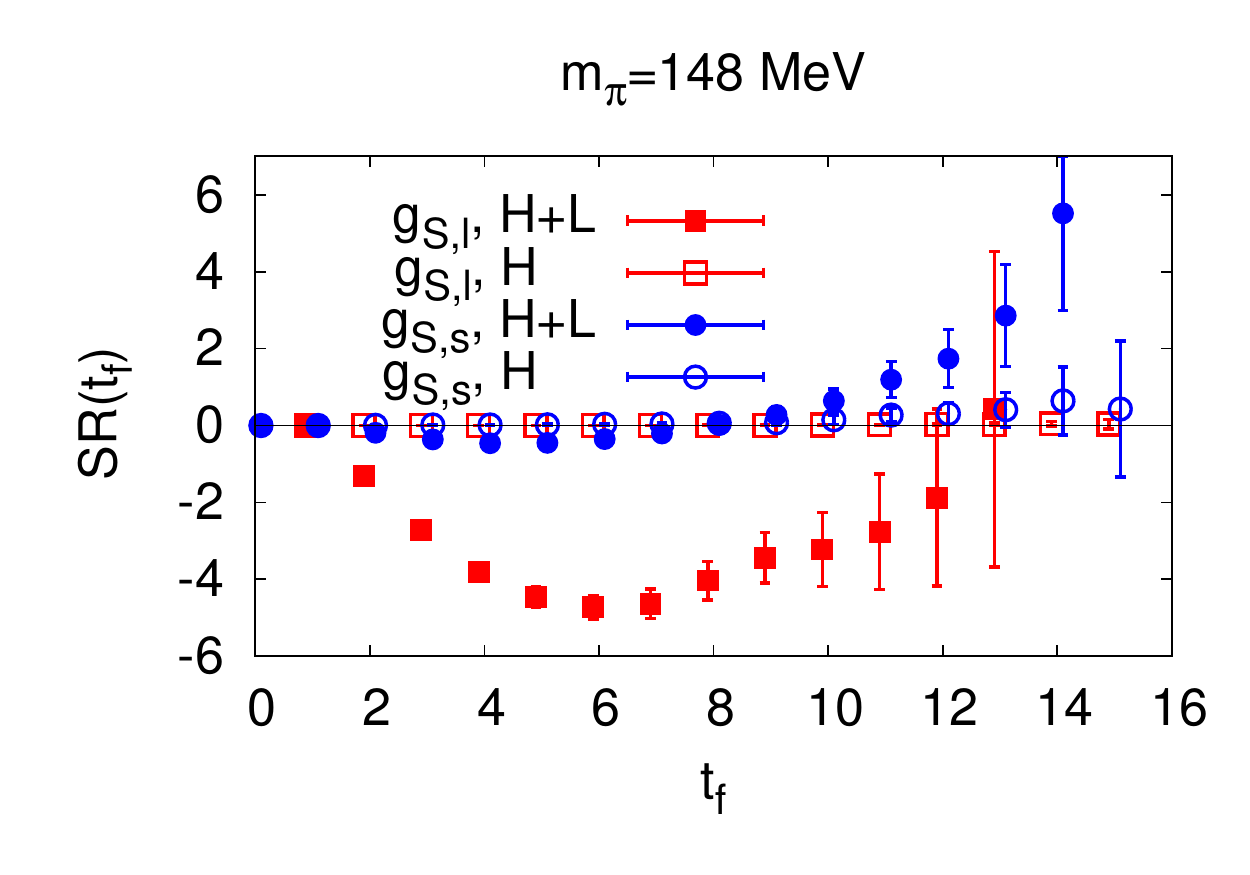} 
  \includegraphics*[scale=0.7]{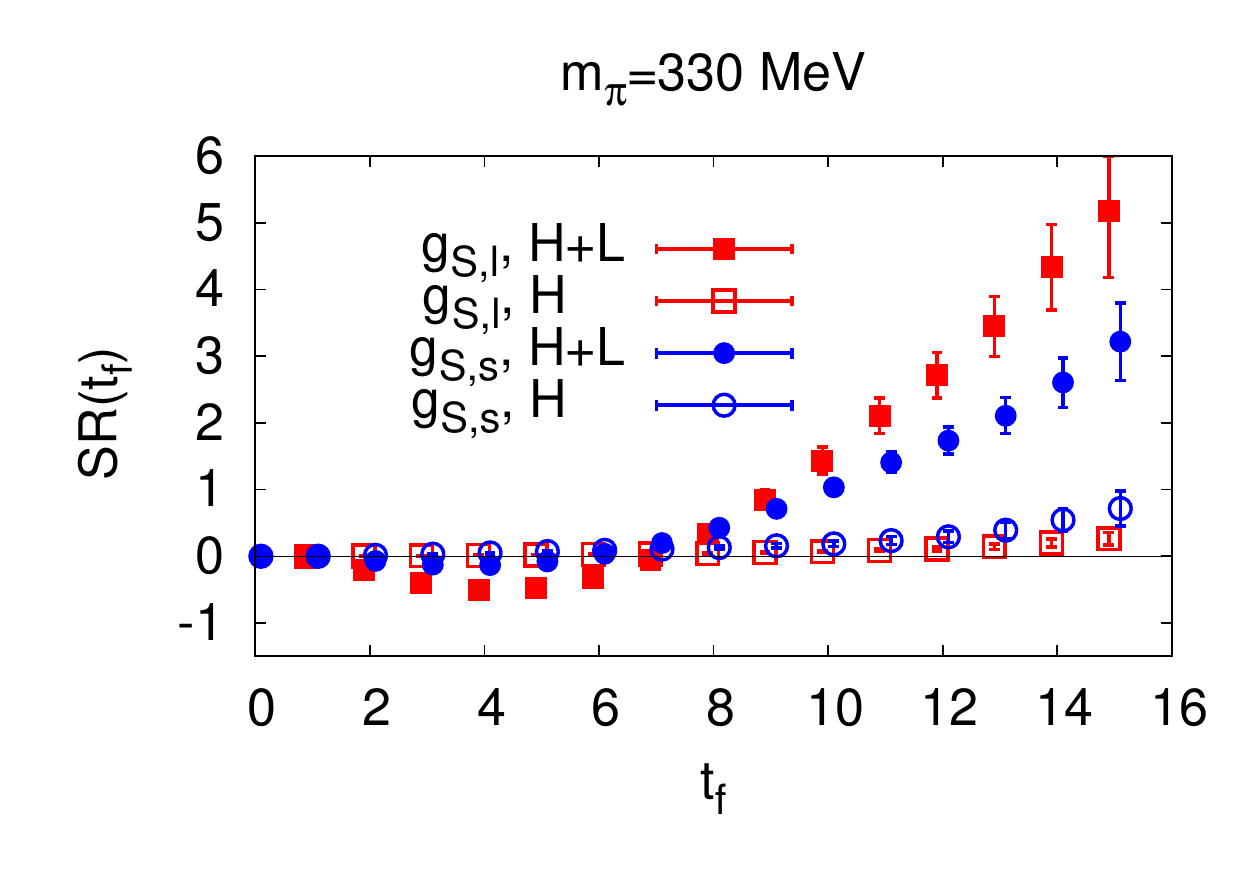} 
 \caption{The light ($m_{\pi}$=148/330 MeV in the upper/lower panel respectively) $g_{S,l}$ and strange quark
loops $g_{S,s}$, with the quark mass in the nucleon the same as that of
the light quark loop, from the
48I lattice are plotted. High-mode contribution (H) and the sum of the high- and
low-mode contributions (H+L) to the DI part of the scalar matrix
elements are shown separately. The contributions from the stochastic high-mode part of the
quark loops are quite small and the results are dominated by the exact low-mode part of the quark loop, so that the LMA method is very effective.}\label{fig:hm3}
\end{figure}

The fact that the long-distance part of the proton two-point correlation function is dominated by the precise low-lying eigensystem of $D_c$ allows us to use a larger residual of $10^{-4}$ for the high-mode inversion without affecting the final accuracy. We also used the low-precision inversion with the same residual for the quark loops, since most of the contribution to the disconnected insertion part of $g_S$ comes from the low-mode part of the quark loop, as shown in Fig.~\ref{fig:hm3}. So we can treat the quark loop with the scalar insertion as almost being exact. Note that we need 4 inversions to get a set of the noise propagators, two for different time slices and two for odd-even dilution for the spatial grid~\cite{Gong:2013vja}.

On  the 48I ensemble, the cost of a low-precision (with residual $10^{-4}$) inversion is just 1/3 of that for a high precision inversion (with residual $10^{-7}$) and the final cost is equivalent to 37 high-precision inversions per configuration, 32 for the proton propagators with the overhead of LMS, and 5 for the 4 sets of the noise propagators for the loops..
 The total number of the measurements of the proton propagator in the ideal case is $4^3\time96$ per configuration. Therefore the total number of the measurements of our DI results can be as large as 497,664 in total, if LMS is perfect and all the measurements are independent. 
 
\subsection{Two-state fit}

\begin{figure}[h]
  \includegraphics[scale=0.7]{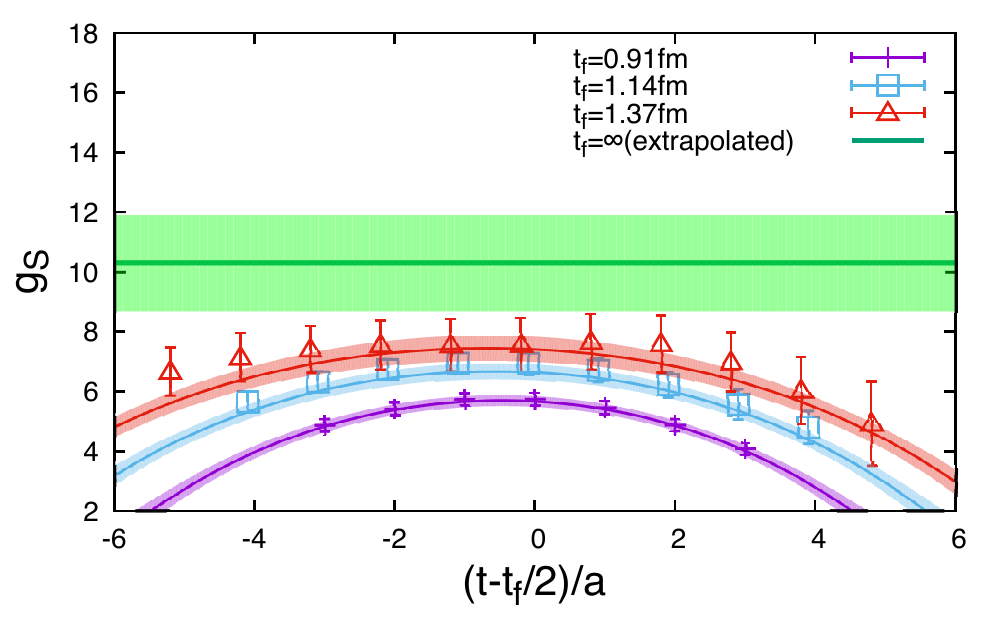} 
  \includegraphics[scale=0.7]{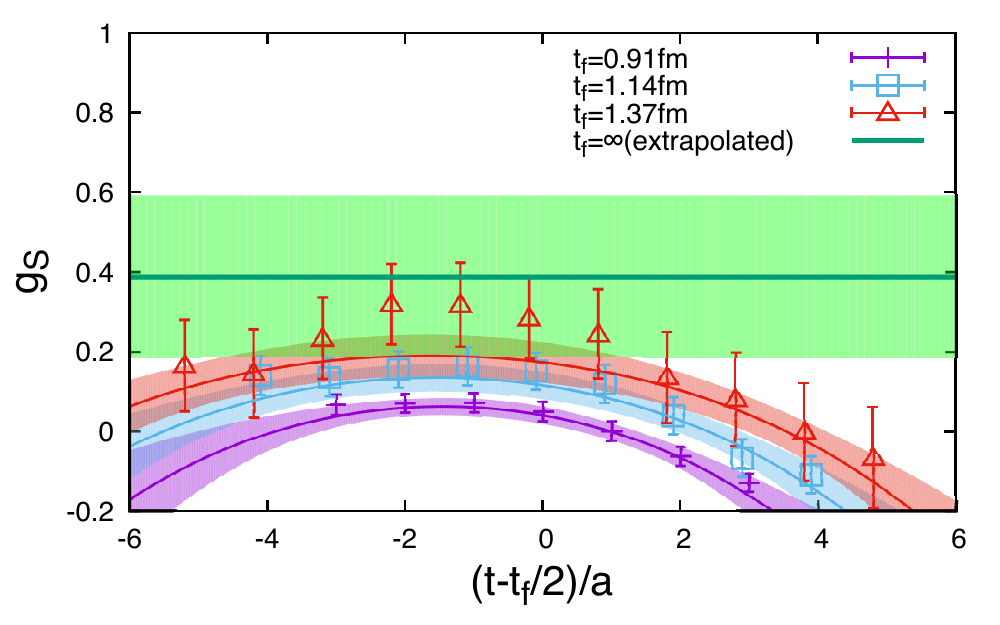} 
 \caption{The ratio $R(t_f,t)$, as a function of the separation $t_f$ (three curves for three separations) and the current position $t$ (the data points on the curves), for each of the light (upper panel) and strange (lower panel) scalar matrix elements in proton, $g_S$, is plotted at $m_{\pi}=148$ MeV (on the 48I ensemble) which is close to the physical point. The green bands show the results extrapolated to infinite separation which correspond to the predictions of $g_S$. The excited-state contaminations are obvious with the final uncertainties much larger than those for the finite separations.}\label{fig:two_term}
 \end{figure}

 To exclude some excited-state contamination, we dropped the data for which the distance between the current insertion and source (or sink) is smaller than 0.2 fm, and applied the two-state fit in Eq.~(\ref{eq:two_term}) to obtain the scalar matrix element in the proton for the light quark and also for the strange quark. We show the case of $m_{\pi}$=148 MeV on the 48I ensemble in Fig.~\ref{fig:two_term}, in which the connected and disconnected insertion parts of the light quark are summed before applying the fit. Note that the curves in Fig.~\ref{fig:two_term} predicted by the fit agree with the data well and their asymmetry  around zero on the horizontal axis is
due to the different treatment of source and sink (smeared source and point sink). 
 
\begin{figure*}[!h]
  \includegraphics[scale=0.8]{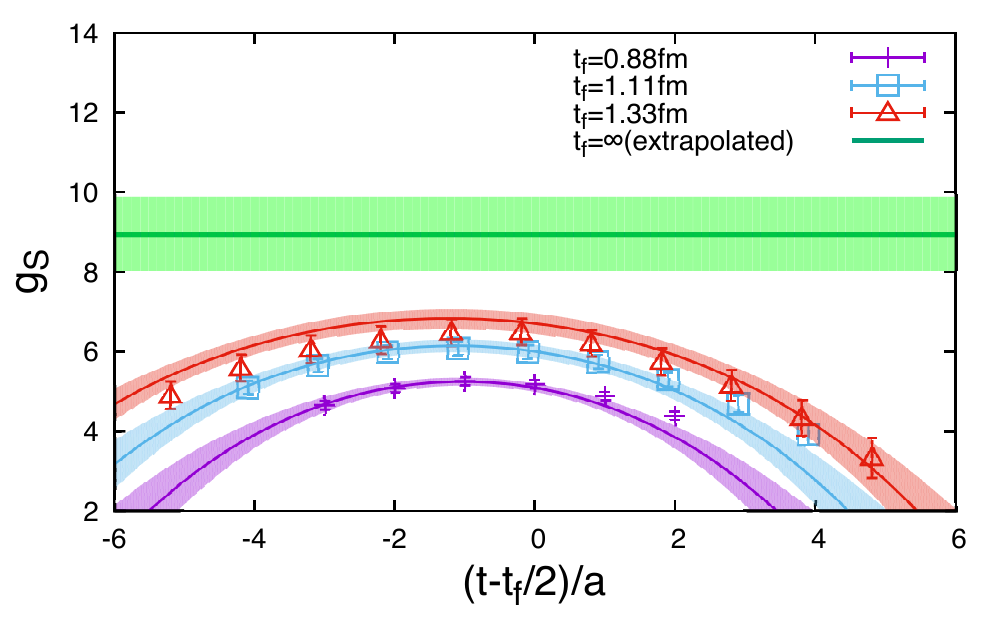} 
\includegraphics[scale=0.8]{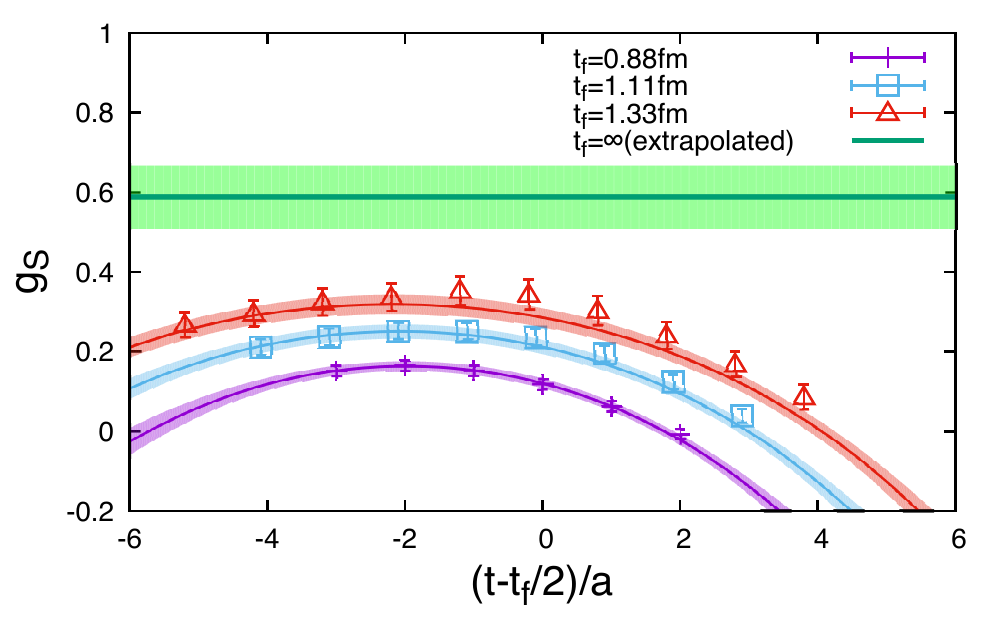} \\
        \includegraphics[scale=0.8]{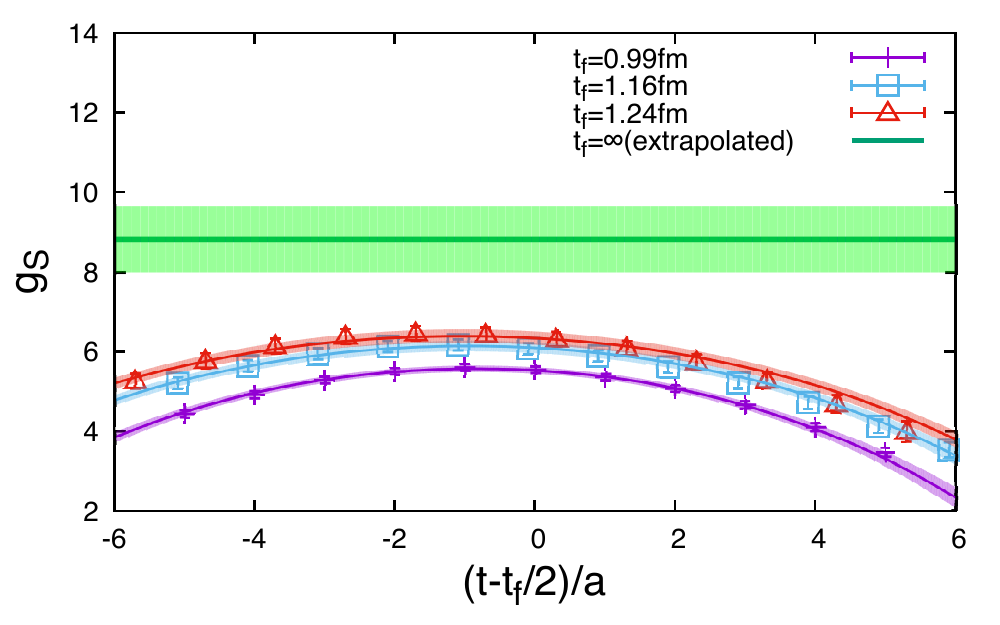} 
   \includegraphics[scale=0.8]{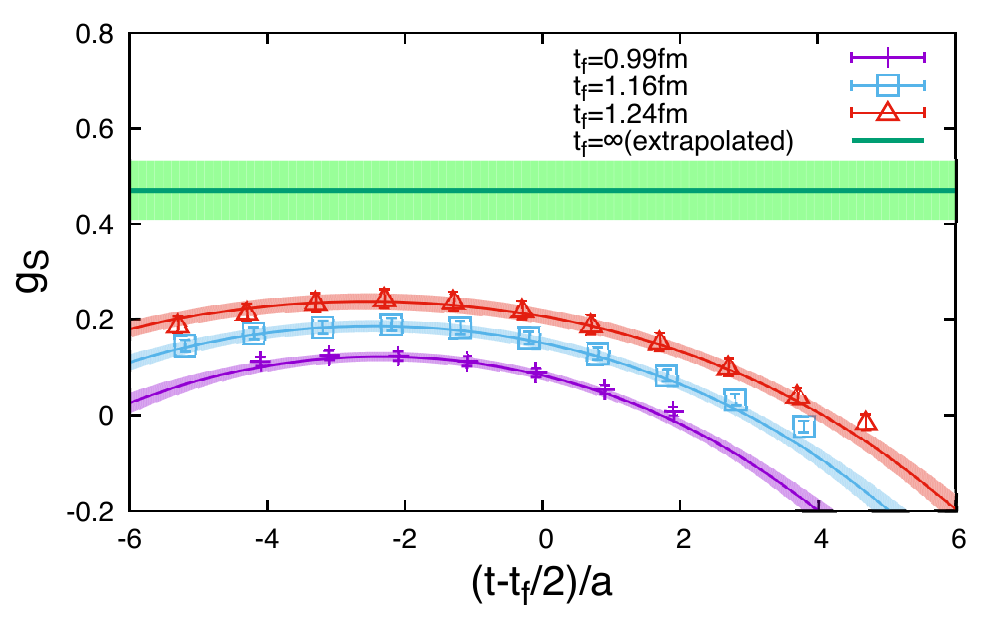} \\
 \includegraphics[scale=0.8]{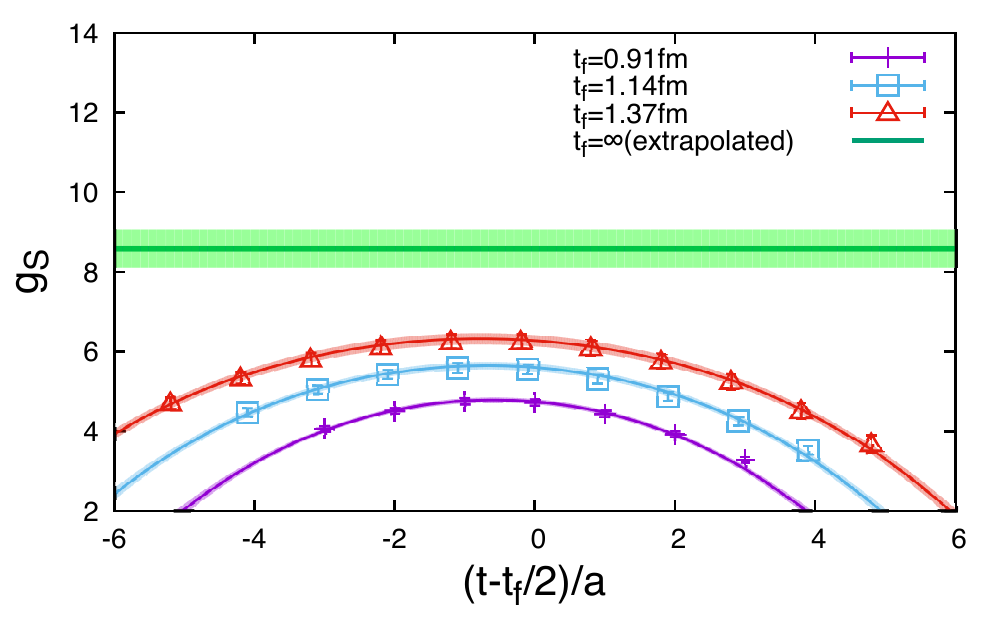} 
    \includegraphics[scale=0.8]{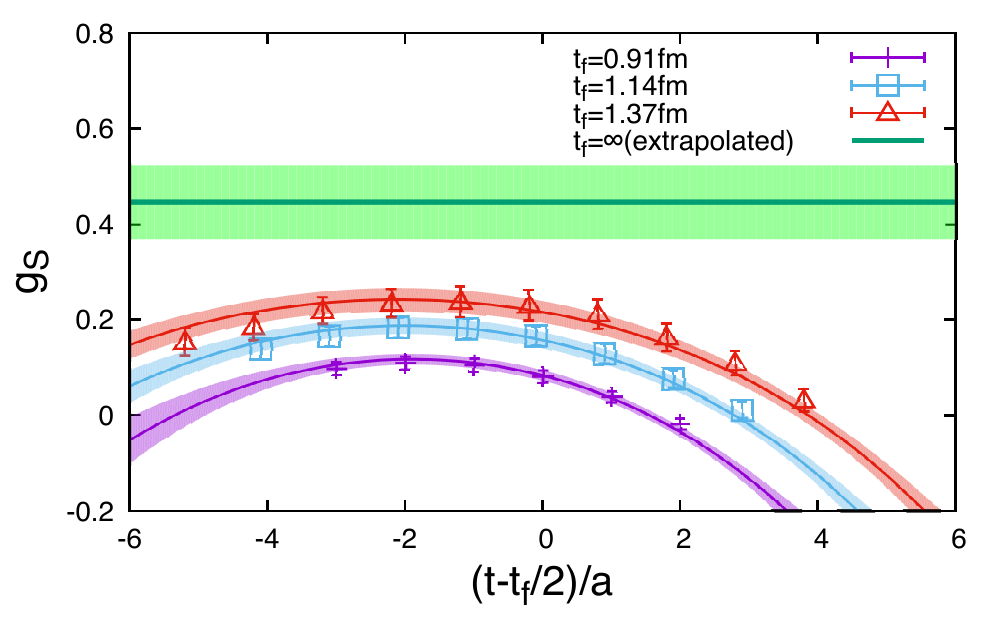} 
  \caption{The ratio $R(t_f,t)$, as a function of the sink-source separation $t_f$ (three curves for three separations) and the current position $t$ (the data points on the curves) for the light and strange matrix elements (left and right panels respectively). Three sets of the panels from top to bottom show the cases for \mbox{$m_{\pi}\sim$ 280 MeV} on the 24I/32I/48I ensembles respectively. The green bands show the results extrapolated to infinite separation which corresponds to the prediction of $g_S$. The excited-state contaminations are obvious and the final uncertainties are larger than those on the finite separations.
 }\label{fig:all}
\end{figure*}

In Fig.~\ref{fig:all}, the ratio $R(t_f,t)$ for $m_{\pi}\sim$ 280 MeV for each of the 24I/32I/48I ensembles is plotted to show the SNR in the relatively heavier pion mass region, and to highlight the qualities of our two-state fit. All curves predicted by the fit agree with the data well, and $\chi^2$/d.o.f is smaller than 1.4 and the $Q$-value is larger than 0.1, for all the quark masses on all ensembles used in this work.

From the fit, we see  that the excited-state contaminations are substantial and the final prediction of $g_S$ (the green band) is one or two sigma higher than the ratio $R(t_f,t)$ with the largest separation. The error bar on $g_S$ is larger than that on  $R(t_f,t)$ at finite separation time $t_f$ due to the extrapolation to infinite $t_f$.

\section{Results}\label{sec:res}

\begin{figure}
  \includegraphics*[scale=0.8]{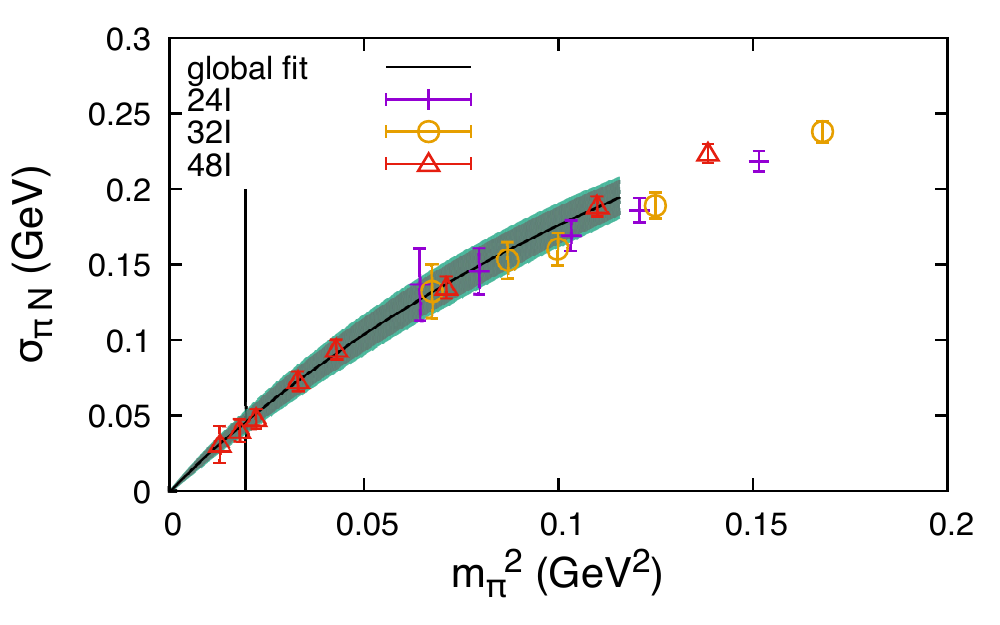} 
  \includegraphics*[scale=0.8]{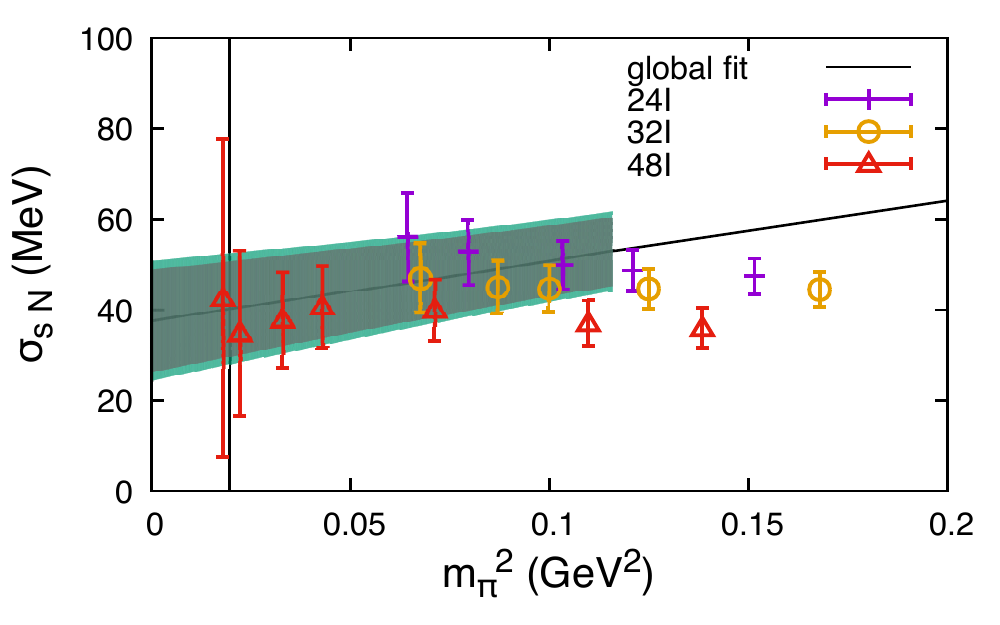} 
 \caption{The summary figures of the light/strange quark content at 18 quark masses on the three ensembles (24I/32I/48I as defined in Table~\ref{table:r0}), as {a} function of the square of the pion mass. Both the lattice spacing and sea quark mass dependence are mild. The curve in each figure shows the behavior in the infinite volume and continuum limits without the partially quenching effect. In each case, the band of the total error is almost the same as  that of the statistical error, and thus is barely visible.}\label{fig:hm1}
\end{figure}

Fig.~\ref{fig:hm1} shows the computed $\sigma_{\pi N}$ and $\sigma_{sN}$ data points for the three ensembles, as a function of $m_{\pi}^2$ corresponding to the valence quark mass. 

The chiral {behavior} of $\sigma_{\pi N}$ as a function of $m_{\pi}$ can be deduced from the chiral behavior of the nucleon mass itself (as suggested by partially quenched SU(2) $\chi PT$~\cite{Beane:2002vq,Tiburzi:2005is,WalkerLoud:2008bp}), by taking the derivative with respect to both valence and sea quark masses,
\begin{eqnarray}
\sigma_{\pi N}(m^v_l,m^s_l,a,L)&=&C^\pi_0m^2_{\pi,vv}+C^\pi_1m^3_{\pi,{vv}}\nonumber\\
&&\!\!\!\!\!\!+C^\pi_2 m^2_{\pi,vs}m^{\textrm{mix}}_{\pi,vs} +C^\pi_3 a^2\nonumber\\
&&\!\!\!\!\!\!+C^\pi_4 (\frac{m_{\pi,vv}^2}{L}-m_{\pi,{vv}}^3) e^{-m_{\pi,{vv}}L},
\end{eqnarray}
with lattice spacing $a$ and lattice size $L$ dependence. The symbol $m_{\pi,vv}$ appearing in the above equation is the valence-valence pion mass and $m^{\textrm{mix}}_{\pi,vs}=\sqrt{m^2_{\pi,vs}+a^2\Delta_{\textrm{mix}}}$  is the mixed valence-sea pion mass. (The value of $\Delta_{\textrm{mix}}$ in our case is small and contributes a shift of only $\sim$10 MeV 
to the pion mass at 300 MeV for the 32I lattice~\cite{Lujan:2012wg}.)

The chiral log term is dropped since it can be fully absorbed by the polynomial terms within our present data precision, and will be considered as a systematic uncertainty. Even for the fit of the proton mass itself where a higher precision is attainable, the coefficient of the chiral-log term obtained by Ref.~\cite{WalkerLoud:2008bp} is still consistent with zero with large uncertainty. The functional form of the volume dependence is derived from the leading order {of} the proton mass~\cite{AliKhan:2003ack,Beane:2004tw} in $\chi$PT. 

For $\sigma_{sN}$, we used the same functional form for the chiral behavior as in
Ref.~\cite{Gong:2013vja} and added a volume-dependent term
\begin{eqnarray}
\sigma_{s N}(m^v_l,m^s_l,a,L)&=&C^s_0+C^s_1 m^2_{\pi,vv}+C^s_2 m^2_{\pi,vs}+C^s_3 a^2\nonumber\\
&&+C^s_4 e^{-m_{\pi,vv}L}.
\end{eqnarray}

We fit all the data points of $\sigma_{\pi N}$ and $\sigma_{sN}$ with \mbox{$m_{\pi}<350$ MeV} simultaneously with a correlated fit, with 1000 bootstrap re-samples on each ensemble, and the final $\chi^2/d.o.f.$ is 0.89 with 16 degrees of freedom. The values of the parameters are summarized in Table~\ref{tab:para}. The curves in the infinite volume and continuum limit without the partial quenching effect are plotted in Fig.~\ref{fig:hm1}, with bands corresponding to the total error. All the data points stay on that curve within one or two standard deviations, which means that the finite lattice spacing, sea quark mass  and volume dependences are mild.

\begin{table}[htbp]
\begin{center}
\caption{\label{tab:para} The fitted parameters for $\sigma_{\pi N}$ and $\sigma_{sN}$.
All the parameters are in units of a power of GeV.}
\begin{tabular}{ccccccc}
\hline
$\sigma_{\pi N}$&$C^\pi_0$ &  $C^\pi_1$ & $C^\pi_2$ & $C^\pi_3$ &  $C^\pi_4$ \\
-- & 2.9(5) & -3.3(1.5) & -0.2(7) & -0.00(3) & 47(111)\\
\hline
$\sigma_{sN}$&$C^s_0$ &  $C^s_1$ & $C^s_2$ & $C^s_3$ &  $C^s_4$ \\
--  & 0.037(13) & 0.00(2) &0.13(6) &-0.02(3) & -19(138) \\
\hline
\end{tabular}
\end{center}
\end{table}

\begin{figure*}
  \includegraphics*[scale=0.7]{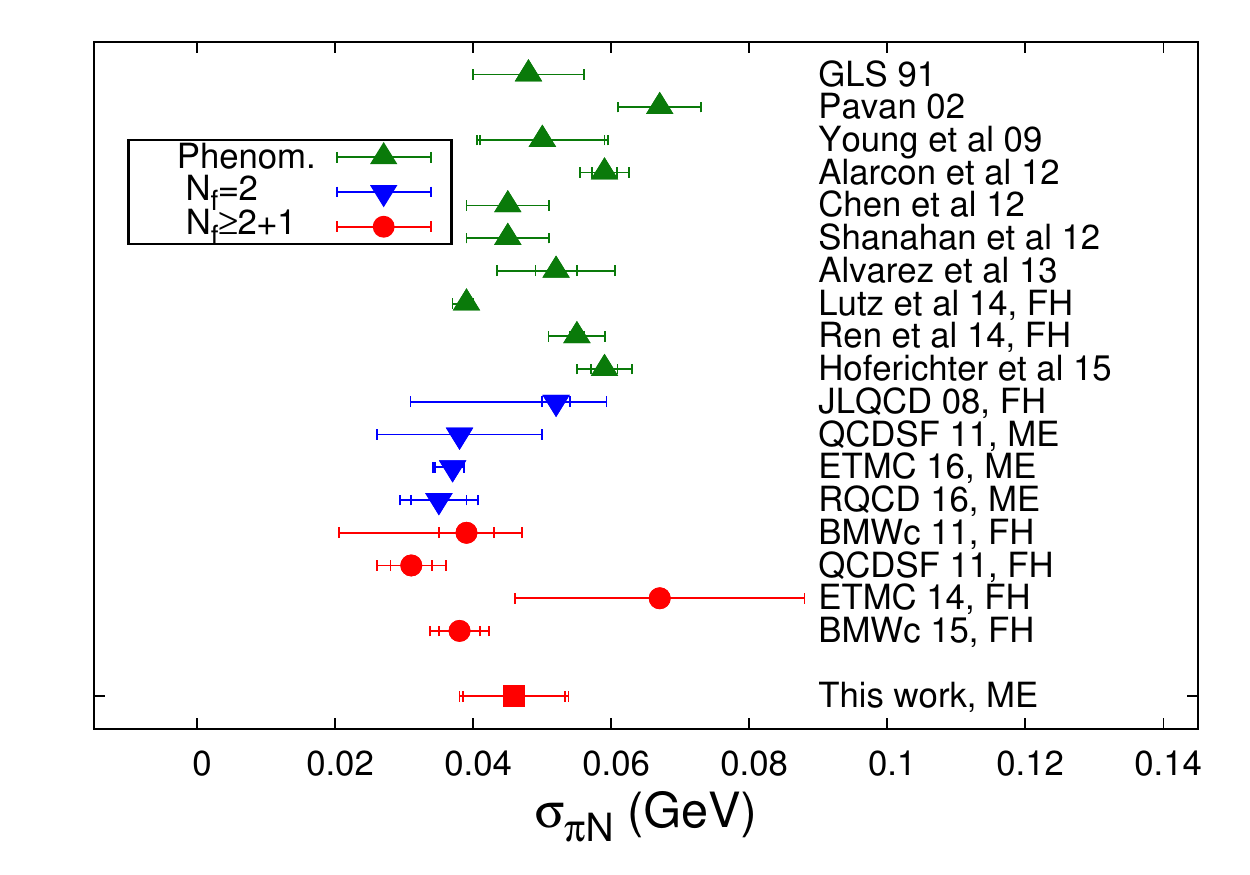} 
  \includegraphics*[scale=0.7]{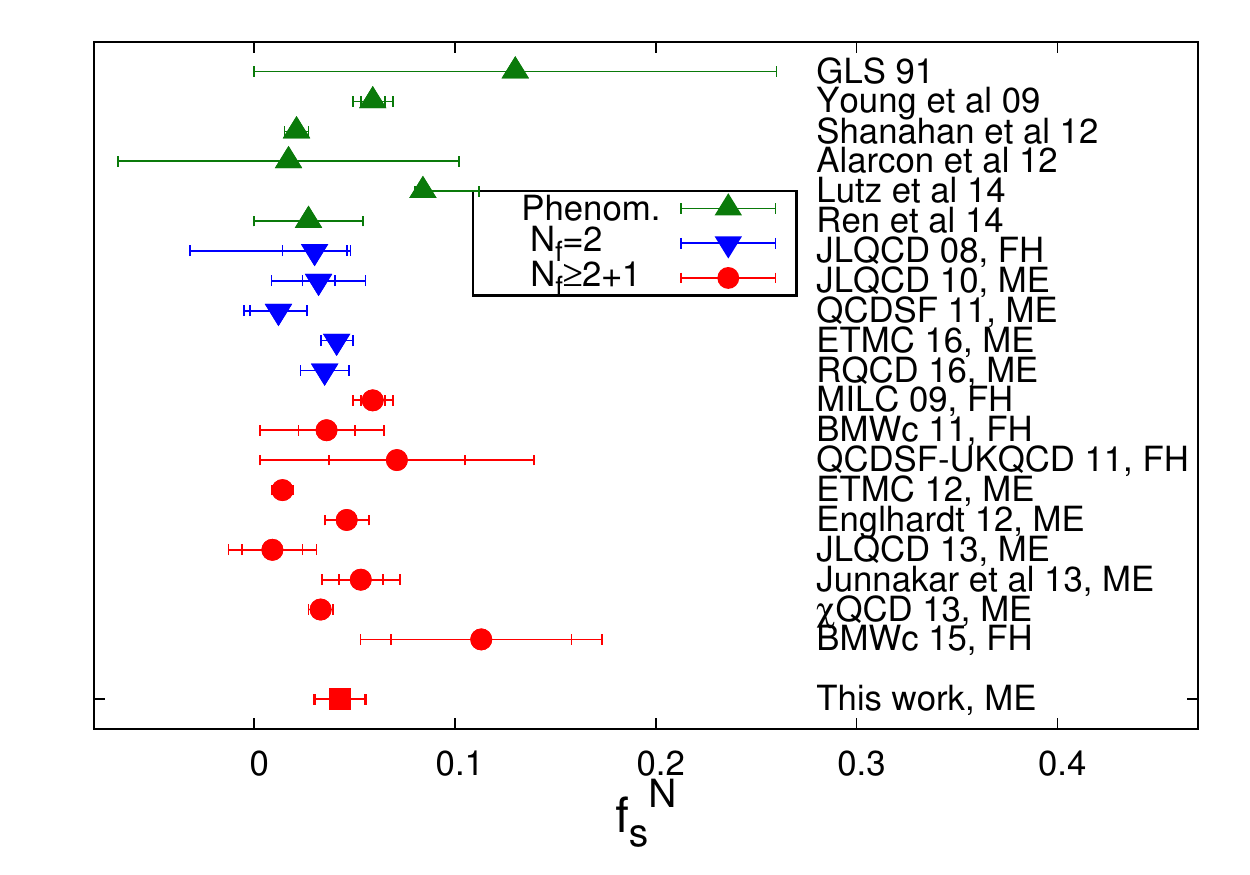} 
 \caption{The results of $\sigma_{\pi N}$ and $f^N_s$, from both phenomenology and lattice simulations. 
 Numbers are from \cite{Gasser:1990ce} (GLS), \cite{Pavan:2001wz} (Pavan), \cite{Young:2009zb} (Young et al.), \cite{Alarcon:2011zs, Alarcon:2012nr} (Alarcon et al.), \cite{Chen:2012nx} (Chen et al.), \cite{Shanahan:2012wh} (Shanahan et al.), \cite{Alvarez-Ruso:2013fza} (Alvarez et al.), \cite{Lutz:2014oxa} (Lutz et al.), \cite{Ren:2014vea} (Ren et al.), \cite{Hoferichter:2015dsa} (Hoferichter et al.),  \cite{Ohki:2008ge,Takeda:2010cw,Oksuzian:2012rzb} (JLQCD), \cite{Bali:2011ks} (QCDSF), \cite{Abdel-Rehim:2016won,Alexandrou:2013xon,Dinter:2012tt} (ETMC), \cite{Bali:2016lvx} (RQCD),  \cite{Durr:2011mp,Durr:2015dna} (BMWc),  \cite{Toussaint:2009pz} (MILC), \cite{Horsley:2011wr} (QCDSF-UKQCD),  \cite{Engelhardt:2012gd} (Engelhardt et al.), \cite{Junnarkar:2013ac} (Junarkar et al.), and~\cite{Gong:2013vja} ($\chi$QCD).  
 The narrow error bar for each data point is the statistical and the  broad one is that for the total uncertainty. The physical proton mass 938MeV is used to obtain $f^N_s$ in this work.}\label{fig:hm2}
\end{figure*}

We estimate the systematic errors of $\sigma_{\pi N}$ and $\sigma_{s N}$ {as follows:}

Discretization errors:\ \ We estimate the systematic errors by the differences between the fitting predictions in the continuum limit and those from the ensemble with the smallest lattice spacing (32I).

Finite volume corrections:\ \ Similarly, we estimate the systematic errors by the difference between the fitting predictions on the ensemble with the largest volume (48I) and those in the infinite volume limit. 

Chiral extrapolation:\ \ The difference between the fitting predictions at the physical pion mass of the 48I ensemble, and those from the interpolations of the neighboring quark masses are considered as systematic errors. 
 
Strange quark mass:\ \ The strange quark mass we used is 101(3)(6) MeV. Since the scalar element will be smaller when the corresponding quark mass is larger, there is just a 1.0 MeV deviation if we change the strange quark mass by $1\sigma$.
 
Mixed action:\ \ We removed the $\Delta_{mix}$ term in the mixed valence-sea pion mass $m^{\textrm{mix}}_{\pi,vs}=\sqrt{m^2_{\pi,vs}+a^2\Delta_{\textrm{mix}}}$ and repeated the fit to simulate the case with the same action for both the valence and sea quark, and the difference turns out to be two orders of magnitude smaller than the statistical error.

Chiral log:\ \ We added the chiral-log term and repeated the fit for $\sigma_{\pi N}$. The coefficient of the chiral-log term is consistent with zero while the uncertainty of the final prediction increases. The prediction will be changed by 2.2 MeV and we take this as a systematic uncertainty of $\sigma_{\pi N}$.

The final prediction of $\sigma_{\pi N}$ is 45.9(7.4)(2.8) MeV where the first error is statistical and the second systematic, as combined in quadrature from those of  the continuum and volume extrapolations, chiral and strange quark mass interpolations, the use of the mixed action and dropping the chiral log term. That of $\sigma_{s N}$ is 40.2(11.7)(3.5) MeV. We determine that the disconnected insertion part contributes 20(12)(4)\% of $\sigma_{\pi N}$. We compare our results with other lattice determinations and phenomenological results in Fig.~\ref{fig:hm2}.

\section{Summary}\label{sec:sum}

$Summary -$ We have computed $\sigma_{\pi N}$ and $\sigma_{sN}$ {for} 18 quark {masses}  including the physical point {on} three 2+1 flavor ensembles including one with the physical pion mass. Since we use chiral fermion in this calculation, there is no additive renormalization of the quark mass for the valence overlap fermion and $\sigma_{\pi N}$ and $\sigma_{s N}$ are renormalization group invariant. As a result, there should be no concern about flavor-mixing of the scalar matrix elements. A global fit is employed to take into account chiral interpolation, finite lattice spacing, and finite volume effects.
The total uncertainty for $\sigma_{\pi N}$ we achieved {is} 17\%. 
Our result straddlies those of the lattice simulations with Wilson-type fermions and the phenomenological predictions, while none of them can be excluded by our present uncertainty. More precise measurements for the disconnected insertion part are required to make a clear adjudication.

The error of $\sigma_{s N}$ is somewhat larger than the former estimate of our collaboration (40(12) MeV versus 33(6) MeV)~\cite{Gong:2013vja}, mostly due to a better control of the excited-state contamination. Even so, it is still the most precise result among 2+1 flavor lattice calculations today which include all the systematic uncertainties. Our results show that the contributions from the quark mass of the two light flavors  and that from the strange flavor are close to each other. Based on the values of $\sigma_{\pi N}$ and $\sigma_{sN}$, we obtain the ratio $y=0.09(3)(1)$.

\section*{Acknowledgments}

We thank the RBC and UKQCD collaborations for providing us their DWF gauge configurations. This work is supported in part by the U.S. DOE Grant No.\ DE-SC0013065. A.A. is supported in part by the National Science Foundation CAREER grant PHY- 1151648 and by U.S. DOE Grant No. DE-FG02-95ER40907. \mbox{Y. Y.} also thanks the Department of Energy's Institute for Nuclear Theory at the University of Washington for its partial support and hospitality. This research used resources of the Oak Ridge Leadership Computing Facility at the Oak Ridge National Laboratory, which is supported by the Office of Science of the U.S. Department of Energy under Contract No. DE-AC05-00OR22725.

\bibliographystyle{apsrev4-1}
\bibliography{reference.bib}

\end{document}